\documentclass[aps,pra,preprint,showpacs]{revtex4}

\usepackage{graphicx}
\usepackage{dcolumn}
\usepackage{bm}
\usepackage[normalem]{ulem}

\usepackage{color}

\begin{document}
\title{A continuous model for bosonic hard spheres in     
quasi one-dimensional optical lattices}

\author{C. Carbonell-Coronado}
\author{F. De Soto}
\author{M.C. Gordillo}
\affiliation{Departamento de Sistemas F{\'i}sicos, Qu{\'i}micos 
y Naturales. Facultad de Ciencias Experimentales. Universidad Pablo de
Olavide, Carretera de Utrera, km 1, E-41013 Sevilla, Spain}

\date{\today}

\begin{abstract}
By means of diffusion Monte Carlo calculations, we investigated the quantum 
phase transition between a superfluid and a Mott insulator  
for a system of hard-sphere bosons in a quasi one-dimensional 
optical lattice. For this continuous hamiltonian, we  
studied how the stability limits of the Mott
phase changed with the optical lattice depth and the transverse confinement
width. A comparison of these results to those of a one-dimensional 
homogeneous Bose-Hubbard model indicates that this last model describes accurately the phase diagram only in the 
limit of deep lattices. For shallow ones, our results are comparable to those of the sine-Gordon model in its limit
of application.  
We provide an estimate of the critical parameters when none of those models are realistic descriptions
of a quasi one-dimensional optical lattice.    
 
\end{abstract}


\maketitle

\section{INTRODUCTION}

When a pair of laser beams interfere to produce a standing wave, 
the net effect is the generation of a periodic potential in whose minima alkali metal 
atoms can be trapped \cite{bloch1,bloch2,greiner}. If we use more than 
a single pair of laser beams, those potential minima  
can be regularly located to form a three dimensional pattern called ``optical lattice''.         
The defining parameters of these structures are the potential depth,
$V_0$, the distance between potential minima, $\lambda/2$, and their
distribution in space. 
Here, $\lambda$
is the wavelength of the laser and determines another relevant 
quantity, the recoil energy, $E_R = h^2/2m\lambda^2$ ($h$, Planck constant; $m$, mass of the atom).    
Optical lattices of almost any geometry can be built \cite{bloch3}, 
but quasi one-dimensional (quasi 1D) ones have been experimentally favored \cite{paredes,haller,stoferle,clement,kruger,fabri} 
in the hopes of obtaining a Tonks-Girardeau (TG) gas \cite{paredes,kinoshita,cazalilla}. 

Almost all the theoretical treatment of optical lattices has been made using as a 
reference the Bose-Hubbard model (BH) \cite{bloch2,cazalilla}. This is a discrete model
whose defining parameters, $J$ and $U$, can be deduced from a more general continuous hamiltonian 
through a series of approximations \cite{jaksch}.  However, in this work, instead of that model, we 
have used continuous hamiltonians as the ones 
already employed to describe neutral atoms loaded in harmonic traps \cite{cazalilla,dun,gir,gregori,blume,gregori2,pollet} 
or homogeneous systems \cite{gregori3,boro}, to represent the quasi one-dimensional optical lattices we are interested in. 
To our knowledge, a similar approach has only been used for full three
dimensional (3D) arrangements \cite{pilati}, and for strictly one-dimensional ones \cite{feli1,feli2}.       

It can be shown \cite{bloch2} that the optical lattice potential 
has the form:
\begin{equation}
V_{ext}(x,y,z)=V_x\sin^2(k_x x)+ V_y \sin^2(k_y y) + V_z \sin^2(k_z z) 
\end{equation}
in which the parameters $V_x,V_y$ and $V_z$, could be different in the three spatial directions.  
To generate a quasi one-dimensional system, the standard practice 
\cite{paredes,haller,stoferle,clement,kruger,fabri} is 
to start with a very high and common value ($V_x = V_y = V_z$ = 20-40$E_R$), and relax one of them afterwards.
The chosen one defines the longitudinal axis of the optical lattice. This means that, for instance, we can  
express the external potential as: 
\begin{equation}
V_{ext}(x,y,z)=V_0\sin^2(k z)+ \frac{1}{2} m \omega_{\perp}^2 (x^2+y^2) \ ,
\label{pot}
\end{equation}
where $\omega_{\perp}$ is the transversal frequency of the confining harmonic potential (we have made $V_z = V_0$),
and $k =  2 \pi /\lambda$. 
The full 3D hamiltonian is then:
\begin{equation}
 H =   \sum_{i=1} ^N  \left[-\frac{\hbar^2}{2m} \nabla^2 + V_{ext} (x_i,y_i,z_i) \right]  
+ \sum_{i<j} V(r_{ij}) \,
\label{3Dh}
\end{equation}
where $N$ is the number of particles of mass $m$ loaded in the quasi 1D optical lattice, and $r_{ij}$ stands for
the distance between each pair. $V(r_{ij})$ is the interparticle potential, for which we used  
a hard spheres (HS) interaction (a common approach). This implies  $V_{int}(r_{ij})=+\infty$ if $r_{ij} < a$ and 0 otherwise. $a$ is
here the diameter of the hard sphere. 
This parameter is sometimes named $a_{3D}$ in the literature \cite{gregori2,gregori3} and models the scattering length of 
the alkali bosons loaded in the optical lattice. In this work, we solved this full 
hamiltonian, not a strictly one-dimensional one. 

Our goal will be to compare the onset of the superfluid-Mott insulator transition (see below) in a pure 1D Bose-Hubbard model and
in the more realistic hamiltonian given by Eq.~(\ref{3Dh}). Since a pure 1D BH model is itself an approximation 
\cite{jaksch}, it is useful to establish numerically the $V_0$ range in which it could be used as a substitute for  
a continuous hamiltonian.
We will also solve the Schr\"odinger equation for the hamiltonian in Eq.~(\ref{3Dh}) for "shallow" (small $V_0$'s) lattices, 
a situation in which the BH model is not longer valid, but where the    
sine-Gordon model \cite{sine} is a good representation for the optical lattice in strictly 1D environments.   
We found that our results were similar to those of the Bose-Hubbard and sine-Gordon models in their
respective limits of high and low $V_0$'s, and can be used to describe the no-man's land in between.    
 
The method to solve the hamiltonian in Eq.~(\ref{3Dh}) will be described in the following section. 
Next, and for the sake of comparison, we will display some results for quasi one dimensional systems with $V_0$=0.
The stability limits of the Mott insulator phase for different optical lattices will be displayed next, together with a 
numerical comparison with those derived from the Bose-Hubbard and sine-Gordon models.    
We will end with some conclusions.    

\section{METHOD}

To solve the Schr\"odinger equation for the hamiltonian given in Eq.~(\ref{3Dh}), we used a Diffusion Monte Carlo (DMC) algorithm. 
This technique allowed us to obtain numerically the ground state of all the boson arrangements we were interested in.  
Those ground states are expected to be reasonable approximations to the experimental systems, due to the extremely low temperatures  
typical of optical lattice studies. 
The numerical solution derived from DMC is exact 
within the statistical uncertainties derived from 
the use of the so called {\em trial function} \cite{boro94}. This is an initial approximation to the ground state wave function
that guides 
the sampling of the phase space. In our case, 
\begin{equation}
\Phi({\bf r_1},\cdots,{\bf r_N}) = \prod_{i=1}^N \psi (x_i,y_i) \prod_{j=1}^N \phi (z_i) \prod_{l<m=1}^N \Psi(r_{lm}) 
\end{equation}
where {\bf r$_i$}$(x_i,y_i,z_i)$ are the positions of each of the $i$ neutral atoms in the optical lattice. 
$\psi (x_i,y_i)$ is the exact solution of the transversal potential. i.e., a Gaussian  
whose variance is related to the harmonic confining frequency 
$\omega_{\perp}$ by $\sigma^2=\frac{\hbar} {m \omega_{\perp}}$.  $\sigma$ is commonly used as a 
measure of the width of the quasi one-dimensional confining ``tube'' \cite{ols,gregori,blume,gregori2,gregori3}. 
Following Ref. \onlinecite{feli1}, $\phi(z_i)$ is 
the ground state wave function of a single particle under a external potential given by $V_{ext}(z)=V_0\sin^2(k z)$, while  
for the remaining part of the {\em trial function} we used  
\begin{equation}\label{inter_trial}                                    
\Psi(r_{ij})=\left\{
\begin{array}{lr}
0 &  r_{ij} < a\\
1-\exp(-\beta(r_{ij}-a)) & r_{ij} \ge a.
\end{array} 
\right. 
\end{equation}
Here, $\beta$ is a variational constant that minimizes the energy per particle at each density and 
$\sigma$.

In our calculations, we considered $\sigma$'s in the range $a$ - 30$a$, $a$ being the diameter of the hard sphere.
Three wavelengths, $\lambda/a=50,100,200$ (given in reduced units) were used. 
Those values of 
$\lambda/a$ are in the range of the experimental ones  for both $^{87}$Rb and $^{133}$Cs \cite{paredes,haller,fol,gem,bakr}. 
To produce the densities displayed in 
the figures below, we used simulation cells of length 20$\lambda/a$ and a variable number of particles, 
ranging from zero to twice the number of the potential minima  
(a maximum of 80 particles in 40 wells). All the energies 
will be given in units of $E_R$, the recoil energy defined above.   

\section{RESULTS}

\subsection{$V_0$ = 0}

\begin{figure}
\begin{center}
\includegraphics[width=8.5cm]{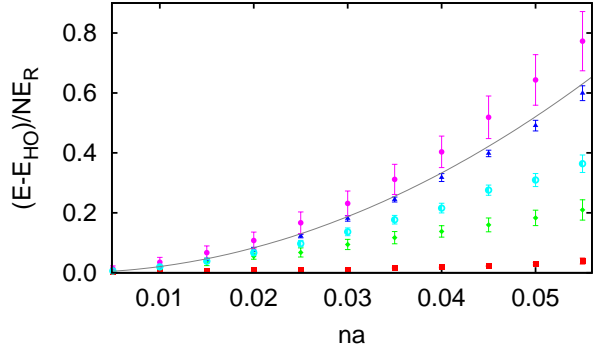}
\caption{
(Color online) Energy per particle (in units of $E_R$ and after subtracting the corresponding harmonic oscillator one) for $V_0/E_R=0$ versus the longitudinal density, 
$na$, for different transversal confinements. Full squares, $\sigma/a=17.78$; diamonds, $\sigma/a=5.62$; lower circles, $\sigma/a =3.16$; triangles, 
$\sigma/a=1.78$; upper circles, $\sigma/a=1$. We show also the result for a 1D TG gas as a full line.
}
\label{noptical}
\end{center}
\end{figure}

In Fig.~(\ref{noptical}), we considered several hard spheres arrangements 
with quasi one-dimensional confinement but no optical lattice in the longitudinal direction ($V_0$ = 0 in Eq.~(\ref{pot})). 
This is a full 3D system, whose energies per particle are displayed after 
the subtraction of the corresponding to the harmonic trap ($E_{HO} = \hbar \omega_{\perp}$). If we go from top to bottom in 
Fig.~(\ref{noptical}) the confinement becomes looser, from $\sigma/a$ = 1 for the upper curve, to $\sigma/a$ =17.78 for the lowest
one. Those results are not strictly comparable to those of Refs. \onlinecite{gregori,blume,gregori2}, since in those works
the 3D hamiltonian includes an additional harmonic confinement in the longitudinal direction (of frequency $\omega_z$), which is 
absent here. However, Ref. \onlinecite{gregori2} gives us a result qualitatively similar to the one displayed in Fig.~(\ref{noptical}):
when $\sigma \sim a$, for a HS interaction and $\omega_z/\omega_{\perp} =$ 0.01, the longitudinal energy per particle is larger 
than the corresponding to a TG 
gas ($E= \pi ^2 \hbar^2 n^2/6m$; $n = N/L$, $L$ length of the simulation cell containing the 
$N$ particles). On the other hand, when $\sigma/a > 1$, the energy per particle is lower than the one for that pure 1D limit. 
This is exactly what is displayed in Fig.~(\ref{noptical}). In addition, we can see that for all confinements the energies 
per particle increase smoothly with $n$, in stark contradiction to what will happen when $V_0/E_R \neq 0$. 
This monotonous increase of the energy with the density is characteristic of a fluid. In any case, we have to bear in mind
that our results are not strictly comparable to those of a TG gas, since this is a purely 1D model.                          
In addition, the size of the particles in a TG model is supposed to be zero, i.e., a TG gas is made of ``hard points'' instead of hard spheres.  

\subsection{Superfluid to Mott Insulator phase transition}

\begin{figure}
\includegraphics[width=8.5cm]{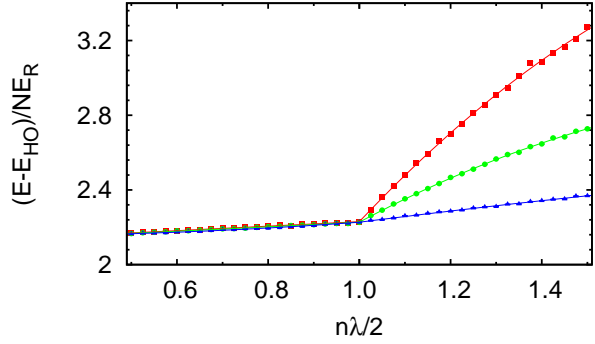}
\caption{(Color online) Same as in previous figure, but for arrangements with $V_0/E_R=6.3$. The density is displayed as the filling 
fraction for different $\left(\sigma/a\right)$ values. Squares, $\sigma/a=1.78$; circles, $\sigma/a=3.16$; triangles, $\sigma/a=5.62$. 
Lines are third order polynomial fits and are intended as guides-to-the eye. Error bars are of the same size  as the symbols 
and were not displayed for simplicity.
}
\label{kink}
\end{figure}

\begin{figure}
\includegraphics[width=8.5cm]{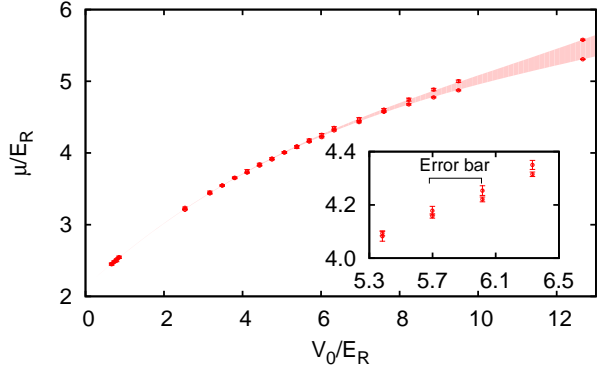}
\caption{(Color online). Values of $\mu/E_R$ calculated using Eq.~(\ref{mudef}) for densities above (upper symbols) and below
(lower symbols) at $n \lambda/2=1$. In this example, $\lambda/a$ = 50 and $\sigma/a \sim 8$. From this set of results, $(V_0/E_R)_C = 5.8 \pm 0.2$.}
\label{muf}
\end{figure}

As an example of the behavior of the energy as a function of the transverse confinement in an optical lattice, we display in   
Fig.~(\ref{kink}) the case for $V_0/E_R = 6.3$ for several values of $\sigma/a$. The density is shown as filling fraction
($n \lambda/2$), the average number of atoms per potential well. In this figure, $\lambda/a$ = 50, but
the trend is common to  many other values of $V_0/E_R$ and $\sigma/a$. In this particular case, we observe that the 
slope of the energy per particle, when it is represented as a function of the density, can change around $n \lambda/2$ = 1. 
This is due to the interaction   
between atoms inside the same potential well. From the energy slope and the energy per particle at density $n$, we can 
obtain the chemical potential, $\mu$, through,
\begin{equation}
\frac{\mu}{E_R} = \frac{\partial \frac{E}{E_R}}{\partial N}
= n \frac{\partial \frac{E}{N E_R}}{\partial n} + \frac{E}{N E_R}.
\label{mudef}
\end{equation}
Eq.~ (\ref{mudef}) implies that when the energy slope changes at a particular density, a discontinuity in the chemical potential is produced.  
Following Refs. \onlinecite{bloch2,feli1,batrou1,batrou2,laz},
we used the appearance of that discontinuity as a signature for the existence of an incompressible ($\Delta n/\Delta \mu$  = 0) Mott insulator phase at  
$n \lambda/2$ = 1 \cite{bloch2}. Not all systems exhibit such discontinuity in $\mu$; $V_0/E_R$ has to reach 
a particular limit before we can see that change. To obtain that critical value, we considered arrangements with different  
$V_0/E_R$, $\sigma/a$, and $\lambda/a$'s, and calculated their chemical potentials just 
below and above $n \lambda/2$ = 1. We did that by performing third order polynomial fits to the energy per particle vs. density
for filling fractions in the ranges [0.9:1] and [1:1.1] and applying Eq.~(\ref{mudef}) at a filling fraction of one.   
We considered both values of $\mu/E_R$ to be identical if their
differences were lower than their respective error bars, obtained from the least squares fits.      
The critical value for the appearance of a Mott insulator phase,  
($V_0/E_R$)$_C$, was calculated as the average between the last $V_0/E_R$ at which the $\mu$'s are identical, and the first one 
at which they are not. An example of this entire procedure is given in Fig.~(\ref{muf}).  
The existence of 
size effects was checked by performing calculations with bigger simulation cells (80 potential wells and $\sim$ 80 atoms), and found  
negligible.  
           
\begin{figure}
\includegraphics[width=8.5cm]{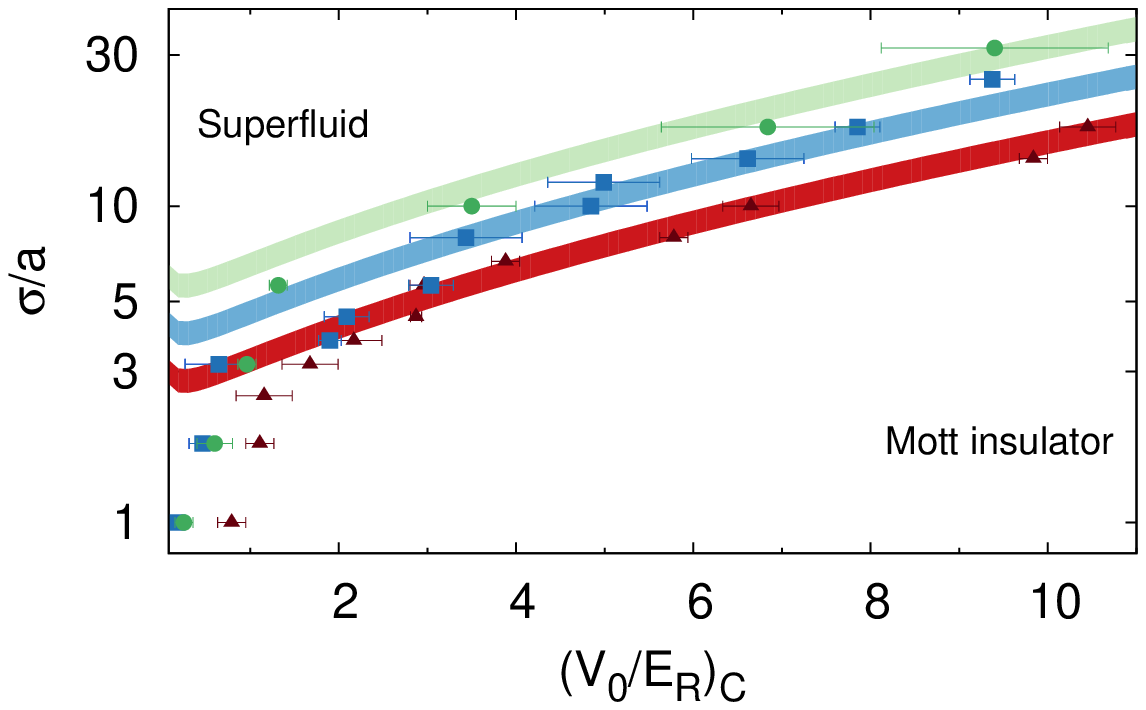}
\caption{(Color online).
Critical values of $\left(V_0/E_R\right)_C$  
for different $\left(\sigma/a\right)$'s. 
Triangles, squares and circles show the simulation results for $\lambda/a=50$, $\lambda/a=100$ and $\lambda/a=200$, respectively.
The three stripes correspond to the $\left(V_0/E_R\right)_C$'s values predicted for the 1D BH model for
each $\lambda/a$ considered. From top to bottom, $\lambda/a=200$, $\lambda/a=100$ and $\lambda/a=50$.     
}
\label{s1}
\end{figure}

Fig.~(\ref{s1}) gives the critical values of $V_0/E_R$ obtained from the above procedure for three different $\lambda/a$'s (50,100 and 200), and a number  
of $\sigma/a$'s. What we see is that for similar $\omega_{\perp}$'s, an increase in $\lambda$ (and 
obviously in the distance between potential wells) reduces the value of $\left(V_0/E_R\right)_C$ necessary to produce a
Mott insulator phase. This is exactly the opposite to what happens in a 3D arrangement, where an increase in $\lambda$ stabilizes the
fluid phase of the atoms loaded in the optical lattice \cite{pilati}. In Fig.~(\ref{s1}) we can observe also another trend: when $\sigma/a$ decreases, 
i.e. when the confinement ``tube'' is thinner, $\left(V_0/E_R\right)_C$ decreases, becoming approximately zero for $\sigma/a$ = 1.
This is in agreement with the results for the strictly 1D system of Ref. \onlinecite{feli2}. There, one can see that 
$\left(V_0/E_R\right)_C$ is exactly zero, i.e., even an infinitesimally small value of $V_0$ is enough to pin the atoms to their
respective potential wells and create a Mott insulator at $n \lambda/2$ = 1. The same conclusion is reached when these systems 
are modeled by a sine-Gordon model \cite{sine}. 

\begin{figure}
\includegraphics[width=8.5cm]{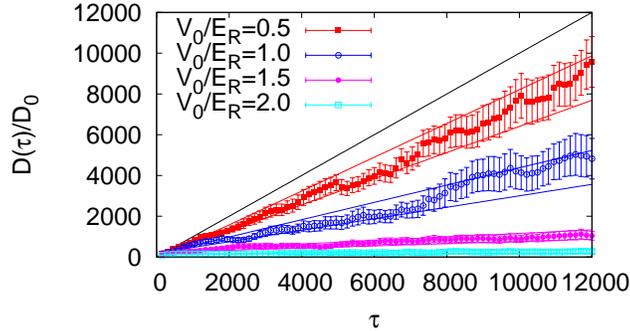}
\caption{(Color online).
Diffusion of the center of mass for $\sigma/a\ = 1.78$ and $\lambda/a=50$  for several $V_0/E_R$ values (in the legend). 
The data have been normalized so that the slope of the curve is the superfluid fraction. The black line corresponding to a 
100\% superfluid fraction has been included for reference. The different bands displayed are the results of performing least squares
linear fits to the respective sets of points defined by the upper and lower limits of the error bars.   
}
\label{sf2}
\end{figure}

The simulation results displayed in Fig.~(\ref{s1}) define two regions for $n \lambda/2$ = 1. In the lower region we have the   
incompressible Mott insulator phase defined above, while the upper part of the figure is labeled ``superfluid''.  
The superfluid fraction of a bosonic system can be extracted from DMC simulations by an extension of
the winding number technique \cite{PhysRevLett.74.1500,PhysRevB.73.224515} as,
\begin{equation}
\frac{\rho_s}{\rho} = \lim_{\tau\to\infty}  \frac{D(\tau)}{ \tau D_0}
\end{equation}
with $D_0=\hbar^2/2m$, and
\begin{equation}
D(\tau) = \frac{1}{2 d N } \langle \left(\vec{r}_{CM}(\tau)-\vec{r}_{CM}(0)\right)^2 \rangle \ .
\label{super}
\end{equation}
where $N$ is the number of particles and $d$ the dimension of the system. $\vec{r}_{CM}(\tau)$ is the position of the center
of mass of all the particles. 
A correct application of Eq.~(\ref{super}) requires that when an atom goes out of the simulation cell, 
one has to keep track of the evolution of the center of mass beyond the simulation cell limits.
$\tau$ is a measure of the number of simulation steps from the point 
in which we start to collect statistics ($\tau = 0$). When each atom is  
located in a particular potential well, the superfluid fraction is zero for
infinitely large simulation times. In Fig.~(\ref{sf2}) we display the results for this estimator in the case $\sigma/a$ = 1.78 and
$\lambda/a= 50$ for different values of $V_0/E_R$. The slopes in          
Fig.~(\ref{sf2}) allow us to obtain superfluid fractions of $0.67 \pm 0.05$ for $V_0/E_R=0.5$, $0.32 \pm 0.07$ for 
$V_0/E_R=1$, $0.06 \pm 0.01$ for $V_0/E_R=1.5$ and $0.02 \pm 0.01$ for $V_0/E_R=2$. Since the $(V_0/E_R)_C$  value for
the same conditions in Fig.~(\ref{s1}) is 1.1 $\pm$ 0.2, this means that at $n \lambda/2= 1$ the onset of the appearance of the Mott insulator
is correlated with a virtual disappearance of the superfluid fraction. The quantum phase transition is then a superfluid-Mott insulator one. 
For densities other than $n \lambda/2 = 1$ the system behaves always as a superfluid. 

\subsection{Comparison with a Bose-Hubbard model}

The Bose-Hubbard model is a {\em de facto} standard to study the physics of atoms in optical lattices, to the extent that many 
experimental data are given in terms of $J$ and $U$, its defining parameters, instead of using the real experimental value, $V_0$.
The hamiltonian of a pure 1D Bose-Hubbard model can be written as \cite{cazalilla},
\begin{equation}
H=-J \sum_{<ij>} b_i^+ b_j+ \frac{U}{2} \sum_i n_i(n_i-1) +  \sum_i \epsilon_i n_i
\label{hub}
\end{equation}
where the $i$'s are discrete sites corresponding to the  minima of the optical lattice potential wells, and  
$<ij>$ stands for pairs of 
nearest neighbors sites. 
$b^+_i(b_i)$ is the creation (annihilation) operator for a boson at the lattice 
site $i$, and $n_i$ the number of atoms at that site. $J$ is the hopping matrix element between nearest-neighbor sites. $U$ is  
the on-site repulsion and $\epsilon_i$ is a term that can consider both the chemical potential and a  possible
longitudinal confinement of frequency $\omega_z$ \cite{jaksch,rigol}. 
Since our hamiltonian does not take into account any longitudinal confinement, 
we can only compare our results to those of 1D {\em homogeneous} BH models   
($\epsilon_i$ = $\epsilon$).    
 
We can transform any 3D hamiltonian describing an optical lattice 
into a BH model by applying the simplifications described in  Ref. \onlinecite{jaksch}. 
This will allow us to  go from the experimental $V_0$, $\omega_{\perp}$ and $\lambda$ parameters, to the derived $U,J$ (and $\epsilon$) ones.  
If, in addition to all the prescriptions of Ref. \onlinecite{jaksch}, we approximate the ground state wave function 
for an atom loaded in a single potential well by a Gaussian, the 
analytical expression for the $J$ parameter \cite{bloch2} is 
\begin{equation}\label{J}
 J\simeq\frac{4}{\sqrt\pi}E_{R}\left(\frac{V_0}{E_R}\right)^{3/4}e^{-2\sqrt{\frac{V_0}{E_R}}},
\label{Jeq}
\end{equation}
This is strictly valid only for  $V_0>>E_R$, even though Eq.~(\ref{Jeq}) is universally used.  
When this condition holds and the interparticle interaction is approximated 
by a pseudopotential, 
the parameter $U$  has the form:
\begin{equation}\label{Ueq}
 U_{1D}=\sqrt{\frac{2}{\pi}}\hbar\omega_{\perp}\left(\frac{V_0}{E_R} \right)^{1/4}\frac{2\pi}{(\lambda/a)}.
\end{equation}
Eq.~(\ref{Ueq}) assumes the point-like interparticle 
interaction of the 1D Lieb-Liniger model ($V(z_{ij})=g \delta(z_{ij})$) \cite{bloch2,cazalilla}.  
From  Eqs.~(\ref{Ueq}) and (\ref{Jeq}), one can obtain an expression that relates the parameters of Eq.~(\ref{3Dh})  
and those of the Bose-Hubbard model: 
\begin{equation}\label{sigma_final}
\frac{\sigma}{a}=\left[  \frac{1}{2\sqrt2\pi} \frac{J}{U} \frac \lambda a \sqrt\frac{E_R}{V_0}  \exp\left(2\sqrt{\frac{V_0}{E_R}}\right)\right]^{1/2}.
\label{s2} 
\end{equation}
By means of this equation, can translate any  $U/J$ value into its corresponding $V_0/E_R(\sigma/a,\lambda/a)$ counterpart and compare 
the onset of the superfluid-Mott insulator transition derived from the 1D BH and the one obtained from our simulations. This is made in
Fig.~(\ref{s1}).        
However, there are several 
$(U/J)_C$ values in the literature \cite{cazalilla,kuh}, all of them
in the range [3.289,4.651]. That is the reason why the Bose-Hubbard model is represented in Fig.~(\ref{s1}) by three different bands 
instead of by 
three simple curves.

We see there that our simulation results, given by the symbols with their corresponding error bars, are in good agreement with the critical 
parameters derived from the BH model when $(V_0/E_R)_C \ge 3$, for all $\sigma/a$'s. This means that all the simplifications described above
make the Bose-Hubbard description unrealistic below that critical value. 
To our knowledge, this is the first time in which a numerical value of $(V_0/E_R)_C$ below which the 1D Bose-Hubbard model
cannot be used to describe a quasi 1D continuous system is given. This is interesting 
since there are experimental setups in which a  $(V_0/E_R)_C <$ 3 is observed \cite{haller}, and it seems to be
no technical reason not to explore that part of the parameter space. 

\section{CONCLUSIONS}

In this paper, we studied the appearance of a Mott insulator phase for a system of hard spheres loaded in quasi one-dimensional 
optical lattices when the filling fraction is exactly one. In other circunstances, we have superfluids. We have then a fully three dimensional system, albeit elongated. 
Our results show that, when $\sigma/a\rightarrow0$, i.e. in the
1D limit, we have a Mott insulator for any  $V_0 \ne $  0. This is in agreement with the results 
of Refs. \onlinecite{feli1} and \onlinecite{feli2} for systems of hard rods in optical lattices similar to the ones
considered here,  
and with the results of the 1D sine-Gordon model \cite{sine} in the limit in which it is comparable to our
model, i.e., when its $\gamma$ parameter tends to  infinity. 
However, when  $\sigma/a >$ 1,  we found that a finite value of $V_0$ is needed to produce 
a Mott insulator. That critical $V_0$ value depends on the degree of transversal confinement and on 
the wavelength of the laser used. For smaller $V_0$'s, the system is a 
superfluid. 

On the other hand, for wider ``tubes'' our results are in good agreement with those of the 1D homogeneous Bose-Hubbard
model with its parameters defined in the standard way. The differences appear only when the optical lattice is shallow enough 
to invalidate the approximations used in deriving the BH hamiltonian. In particular, the BH model is no longer valid when $(V_0/E_R)_C< $3,  
irrespective of the $\lambda/a$ chosen. Below this value,  we provide 
an estimation of the critical value of $V_0$ for the onset of the superfluid-Mott insulator phase transition for different values of 
$\sigma/a$ and $\lambda/a$ for an homogeneous ($\omega_z$ = 0) hamiltonian. This work is then the quasi one-dimensional counterpart of 
Ref. \onlinecite{pilati}, in which the critical parameters for a system of hard spheres loaded in a full three dimensional homogeneous optical 
lattice are calculated. Summarizing, we can say that our simulation results are consistent with those of the BH and sine-Gordon models 
in the parameter range in which those two approximations are expected to be valid. We can also provide estimations of the critical 
$V_0$ values when they are not.     
                      
Even though there are many experimental studies of gases loaded in quasi 1D optical lattices, there are very few from which 
we could obtain the critical parameter for a Mott insulator transition.   
For instance, in Ref. \onlinecite{stoferle}, $\sigma/a \sim$ 10.6, and $\lambda/a \sim$ 160 (we took $a$ = 5.29 nm). This means that we would expect 
a $(V_0/E_R)_C$ value around 4 (an average between those for $\lambda/a$ = 100 and $\lambda/a$ = 200 for the same 
$\sigma/a$ in Fig.~(\ref{s1})). However, the experimental value is in the range 6-8. The same could be said of the results
from Ref. \onlinecite{clement}: for $\sigma/a \sim 10.3$ and  $\lambda/a \sim$ 160, and the same value of $a$, $(U/J)_C$ is in the range 8-10.
Using Eq.~(\ref{s2}), this translates into a $(V_0/E_R)_C$  between 5.8 and 6.7, to be compared to the same value
$(V_0/E_R)_C \sim$  4 from Fig.~(\ref{s1}). 
That would imply that our hard sphere model underestimates the heights of
the potential wells necessary to create a Mott insulator by as much as 50\%. 
One of the reasons for this discrepancy could be that we have not included any 
longitudinal confinement ($\omega_z \ne 0$) in our hamiltonian. Calculations on a pure 1D BH model including that effect have been performed at
finite temperature \cite{rigol}, and indicate that $(U/J)_C$ increases when that effect is included.
On the other hand, at $T=0$ K, (our case), a 1D BH model with $\omega_z \ne 0$ \cite{batro2002}, gave results similar to those of a homogeneous 
model \cite{batrou1}. This suggests that one effect to consider is temperature. In any case, a comparison between the results of 
Ref. \onlinecite{batrou1} and 
Ref. \onlinecite{batro2002} indicates that the study of an homogeneous system is useful, at least, as a reference,
in exactly the same way that the results of Ref. \onlinecite{pilati}
serve as a benchmark for three dimensional systems.   
In the future, the methods presented in this paper 
can be extended to investigate the behavior of cold atoms in optical lattices including a longitudinal confining potential, or
to consider interactomic interactions other than the hard spheres one. 

\begin{acknowledgments}
We acknowledge financial support from the Junta
de Andaluc\'{\i}a Group PAI-205, Grant No. FQM-5987, MICINN 
(Spain) Grant No. FIS2010-18356.
\end{acknowledgments}

\end{document}